\documentclass[twocolumn,twoside]{article}
\usepackage{graphics}
\newcommand{\hb}{\\ \hspace*{2ex}}

\begin{document}
\title{STATISTICAL ANALYSIS OF LARGE-SCALE STRUCTURE OF UNIVERSE}
\author{A.V.\,Tugay\\[2mm] 
\begin{tabular}{l}
 Astronomy and Space Physics Department, Faculty of Physics, \\ Taras Shevchenko National University of Kyiv,\hb
 Glushkova ave., 4, Kyiv, 03127, Ukraine,  {\em tugay.anatoliy@gmail.com}\\
\end{tabular}
}
\date{}
\maketitle

ABSTRACT.

While galaxy cluster catalogs were compiled many decades ago, other structural elements of cosmic web are detected at definite level only in the newest works. For example, extragalactic filaments were described by velocity field and SDSS galaxy distribution during the last years. Large-scale structure of the Universe could be also mapped in the future using ATHENA observations in X-rays and SKA in radio band. Until detailed observations are not available for the most volume of Universe, some integral statistical parameters can be used for its description. Such methods as galaxy correlation function, power spectrum, statistical moments and peak statistics are commonly used with this aim. The parameters of power spectrum and other statistics are important for constraining the models of dark matter, dark energy, inflation and brane cosmology. In the present work we describe the growth of large-scale density fluctuations in one- and three-dimensional case with Fourier harmonics of hydrodynamical parameters. In result we get power-law relation for the matter power spectrum.
\\[3mm]

{\bf 1. Introduction}\\[1mm]

Large-scale structure of the Universe (LSS) is
thought to be a composite of galaxy clusters, superclusters, voids, walls and ﬁlaments. It is very hard to detect extragalactic ﬁlaments with current observational facilities (Tugay, 2014), although future projects such
as ATHENA and SKA (Kale et al., 2016) gives some
perspectives. Attempts to build ﬁlament network on
Sloan Digital Sky Survey data were performed in (Tugay, 2014) and (Chen et al., 2016) using the method
of density ridges. X-ray space observatories gives us
new ﬁndow for LSS studying. First results of X-ray
galaxy distribution based on XMM-Newton observations were get in (Elyiv et al., 2012) and (Tugay, 2012).
Extragalactic ﬁlaments and other LSS details could be
found in future with ATHENA observations (Nandra
et al., 2013; Nevalainen, 2013). This may be also possible due to ATHENA detection and observations of 3.5
keV emission line of sterile neutrino decay (Neronov
\& Malyshev, 2016). Until there are no enough observer galaxy positions to recover internal structure of
single ﬁlaments, the main way of LSS studying is estimation of general statistic parameters of extragalactic
object distribution, such as correlation function, power
spectrum and other. Such parameters should be predicted from cosmological theoretical models including
the theory of growth of primordial density perturbations (Bernardeau et al., 2002).
In this work we consider simple case of gravitational
instability, apply power law assumption for Fourier
harmonics of perturbations and derive a relation for
power spectrum.


{\bf 2. Starting equations}\\[1mm]

LSS formation can be described in Newtonian gravity by the equations of Poisson, Euler and continuity:

\begin{equation}\label{form1}
\Delta \phi =4\pi G\rho
\end{equation}

\begin{equation}\label{form2}
\frac{d(\rho \vec v)}{dt}+(\vec v \cdot \nabla )(\rho \vec v)=-\rho \nabla \phi 
\end{equation}

\begin{equation}\label{form3}
\frac{d\rho }{dt}=-\rho \nabla \cdot \vec v
\end{equation}

We will consider velocity field as irrotational. This allows us introduce velocity potential and divergence:

\begin{equation}
 \overrightarrow{u} =-\nabla \Phi , \theta =\nabla \cdot \overrightarrow{u} 
\end{equation}

Then we can write continuity equation as 

\begin{equation}
 \dot \delta + \theta =0 
\end{equation}

And Euler equation as

\begin{equation}
 \dot {\overrightarrow{u}} + (\overrightarrow{u} \cdot \nabla ) \overrightarrow{u} =-\nabla \Phi 
\end{equation}

or

\begin{equation}
 \dot \theta +\nabla \phi + (\overrightarrow{u} \cdot \nabla ) \overrightarrow{u}=0
\end{equation}

Let's take the divergence from the last equation and use the continuity equation. We will get single equation for continuous matter in own Newton gravity field:

\begin{equation}
 \ddot \delta +H^2\delta +\nabla (\overrightarrow{u} \cdot \nabla ) \overrightarrow{u}=0
\end{equation}

\noindent where H is inverse character time of the system. It may be close by order of value to Hubble constant but not necessary. 
To analyse LSS evolution we will onsider density contrast as superposition of plane waves:

\begin{equation}
 \delta=\sum_{m}\delta _m e^{i(k_mx+\omega _mt)}
\end{equation}

In the linear approximation we can neglect the last term in (8). We will then find the solution for exponential growth of all modes with character time t=1/H. Below we will solve the equation (8) in one-dimentional and three-dimentional weakly non-linear case.\\[6mm]

{\bf 3. One-dimentional case}\\[1mm]

In this case velocity field has single component and the values of $\theta $ and $\Phi $ ane connected with it by simple spatial derivative:

\begin{equation}
 u=-\frac{d\Phi }{dx}, \theta =-\frac{d^2\Phi }{dx^2}
\end{equation}

Now equation (8) can be written as 

\begin{equation}
  \ddot \delta +H^2\delta + \theta ^2 +u\frac{d\theta }{dx}=0
\end{equation}

Now we can write plane wave decompositin for $\Phi $, $u$ and $\theta $:

\begin{equation}
 \Phi=\sum_{m}\Phi _m e^{i(k_mx+\omega _mt)}
\end{equation}

\begin{equation}
 u=\sum_{m}u _m e^{i(k_mx+\omega _mt)}
\end{equation}

\begin{equation}
 \theta=\sum_{m}\theta _m e^{i(k_mx+\omega _mt)}
\end{equation}

Let's substitute last expression to (11) and perform Fourier transformation. We will keep one $\theta $ in the third term and $d\theta /dx$ in the fourth term unchanging. We will get the next expression for harmonics.

\begin{equation}
 \delta _m (H^2-\omega _m^2)+\theta _m \cdot \theta +u_m \cdot \frac{d\theta }{dx}=0
\end{equation}

It can be shown, suprisingly, that in one dimentional case there is simple expression between harmonic amplitudes of $u$, $\theta $ and $d\theta /dx$ such as

\begin{equation}
 \theta _m \cdot \theta =u_m \cdot \frac{d\theta }{dx}
\end{equation}

So we have

\begin{equation}
 \delta _m (H^2-\omega _m^2)+2\sum_{m}\theta _m^2 e^{i(k_mx+\omega _mt)}=0
\end{equation}

This means that $\theta _m=0$ and we can not consider nonlinear dynamics with out assumptions in one dimentional case. (Abell, Corwin \& Olowin, 1989)\\[2mm]

{\bf 4. Three-dimentional case}\\[1mm]

Let's write nonlinear term in (8) in tensor notations

\begin{equation}
 \nabla (\overrightarrow{u} \cdot \nabla ) \overrightarrow{u} =u_{p,q}u_{q,p}+u_qu_{p,pq}
\end{equation}

and change velocity field with scalar potential:

\begin{equation}
 \nabla (\overrightarrow{u} \cdot \nabla ) \overrightarrow{u} =\Phi _{,pq}\Phi _{,qp}+\Phi _{,q}\Phi _{,ppq}
\end{equation}

Then equation (8) takes form

\begin{equation}
  \ddot \delta +H^2\delta +(\nabla \otimes \overrightarrow{u}) \cdot (\nabla \otimes \overrightarrow{u})+\overrightarrow{u} \cdot \nabla \theta =0
\end{equation}

or

\begin{equation}
 \ddot \delta +H^2\delta +\Phi _{,pq}\Phi _{,qp}+\Phi _{,q}\Phi _{,ppq}=0
\end{equation}

Direct tensor product of nabla operator and vector velocity field can be written for plane waves (remember (4)):

\begin{equation}
 \nabla \otimes \overrightarrow{u}=\frac{u_p}{dx_q}=\frac{d}{dx_q}(-ik_p\Phi )=-k_pk_q\Phi
\end{equation}

Double scalar product of such two tensors is equal to 

\begin{equation}
 (\nabla \otimes \overrightarrow{u}) \cdot (\nabla \otimes \overrightarrow{u})=\sum_{p,q}k_p^2k_q^2\Phi ^2=k^4\Phi ^2
\end{equation}

Now let's find the derivatives in the last term in (21) in the plane wave assumption:

\begin{equation}
 \Phi _{,q}=-\overrightarrow{u}=i\overrightarrow{k}\cdot \Phi
\end{equation}

\begin{equation}
 \Phi _{,pp}=\frac{d}{dx_p}(-ik_p\cdot \Phi)=-k^2\Phi=-\theta
\end{equation}

\begin{equation}
 \nabla \theta =-k^2\overrightarrow{u}=k^2\nabla \Phi
\end{equation}

\begin{equation}
 \overrightarrow{u} \cdot \nabla \theta =-i\overrightarrow{k}\Phi \cdot (-k^2\overrightarrow{u})=
  ik^2\overrightarrow{k} \Phi \cdot (-i\overrightarrow{k}) \Phi =k^4\Phi ^2
\end{equation}

Applying all above derivatives, equation (8) takes form

\begin{equation}
 \ddot \delta +H^2\delta +2k^4\Phi ^2=0
\end{equation}

or for Fourier harmonics

\begin{equation}
  \delta _m(H^2-\omega _m^2)+2k_m^4\Phi _m^2=0
\end{equation}

In the opposition to one-dimentional case, now we can get non-trivial result with additional assumptions. First, let's suppose that LSS is periodical in the cube with side equal to L. Then we have straight expression for wavenumber: $k_m=2\pi m/L$. Second, suppose powerlaw relation of harmonics of velocity potential: $\Phi _m=\Phi _0m^{\gamma}$. The last term in (29) must be constant. This gives us

\begin{equation}
 k_m^4\Phi _m^2=(2\pi m/L)^4 \cdot \Phi _0^2m^{2\gamma }
\end{equation}

and finally

\begin{equation}
 \gamma =-2
\end{equation}

So we get inverse power law relation for matter power spectrum which is well agreed with standard cosmological model and analysis of galaxy observations
(Tegmark et al., 2004):

\begin{equation}
 P=P_0(k/k_0)^{-2}
\end{equation}

{\bf 5. Results and conclusion}\\[1mm]

Matter power spectrum parameters were found in this work in simple analytical model. More general case will be considered in the next work under wider assumptions for different functional relations for Fourier harmonics of density and velocity field. \\[2mm]

{\it Acknowledgements.} The author is thankful to V.Zhdanov for the general idea of such study. This research has made use of NASA's Astrophysics Data System. \\[1mm]

{\bf References\\[1mm]}
\noindent\\
Bernardeau F., Colombi S., Gaztanaga E., Scoccimarro R.: 2002, {\it PhR}, {\bf 367}, 1. \\
Chen Y.-C., Ho S., Brinkmann J. et al.: 2016, {\it MNRAS}, {\bf 461}, 3896. \\
Elyiv A., Clerc N., Plionis M.: 2012, {\it A\& A}, {\bf 537}, 131. \\
Kale, R., Dwarakanath, K.S., Vir Lal D. et al.: 2016, {\it JApA}, {\bf 37}, 31. \\
Nandra K., Barret D., Barcons X. et al.: 2013, arXiv:1306.2307. \\
Neronov A., Malyshev D.: 2016, {\it PhRvD}, {\bf 93}, 3518. \\
Nevalainen J.: 2013, {\it AN}, {\bf 334}, 321. \\
Tegmark M., Blanton M.R., Strauss M.A. et al.: 2004, {\it ApJ}, {\bf 606}, 702. \\
Tugay A.V.: 2012, {\it Odessa Astron. Publ.}, {\bf 25}, 142.  \\
Tugay A.V.: 2014, {\it AASP}, {\bf 4}, 42.  \\

\vfill

\end{document}